\documentclass[print]{revtex4}
\usepackage{amsmath,amssymb}
\usepackage{graphicx}

\begin{document}

\title{\large \bf Coupled atomic-molecular condensates in a double-well potential: decaying molecular oscillations}

\author{Hui Jing$^{1,2,3}$\footnote{Electronic address: jinghui@wipm.ac.cn}, Sihong Gu$^{1,3}$ and Mingsheng Zhan$^{1,3}$}

\affiliation{1. State Key Laboratory of Magnetic Resonance and
Atomic and Molecular Physics,\\
 Wuhan Institute of Physics and Mathematics, Chinese Academe of Science, Wuhan 430071, People's Republic of China\\
 2. Department of Chemistry, Hong Kong University of Science and Technology, Kowloon, Hong Kong\\
 3. Center for Cold Atoms, Chinese Academe of Science, People's Republic of China}

\begin{abstract}
We present a four-mode model that describes coherent
photo-association (PA) in a double-well Bose-Einstein condensate,
focusing on the $average$ molecular populations in certain
parameters. Our numerical results predict an interesting
strong-damping effect of molecular oscillations by controlling the
particle tunnellings and PA light strength, which may provide a
promising way for creating a stable molecular condensate via
coherent PA in a magnetic double-well potential.\\

PACS numbers: 03.75.Fi; 03.75.Mn; 05.30.Jp

\end{abstract}

\baselineskip=16pt

\maketitle

\indent The remarkable realizations of Bose-Einstein condensates
(BEC) in cold dilute atomic gases have provided a rich playground
to manipulate and demonstrate various properties of quantum
degenerate gases [1]. Recently rapid advances have been witnessed
for creating a quantum degenerate molecular gase via a magnetic
Feshbach resonance [2-3] or an optical photo-association (PA)
[4-5] in an atomic BEC, and the appealing physical properties of
the formed atom-molecule mixtures were investigated very
extensively under the quasi-homogeneous trapping conditions [2-5].
The coherent PA process not only produces a new species of BEC but
also leads to many interesting quantum statistical effects due to
its nonlinear coupling nature in the dynamics [6]. On the other
hand, the pure atomic condensates in a double-well potential also
attracted considerable interest since many intriguing quantum
phenomena really can appear in this system [7], hence a natural
question arises about the possible new properties of a hybrid
atom-molecule condensate in a double-well potential, by adding an
additional associating light. This problem is also related to that
of molecular formations via PA in an optical lattice [8].

In fact, our previous work in three-mode case already showed that
[9], the coherent PA process in the case of an atomic output
coupler really leads to some novel quantum statistical phenomena,
e.g., the squeezing-free effect for molecules formed by
photo-association of atoms in the propagating mode [9]. In this
paper, we present an effective mean field approach (MFA) to study
the average molecular populations in two wells by numerically
solving the general depleted case beyond short-time limits. We
find that, just for valid MFA parameters range, the new freedoms
of atomic tunnelling and then the formed molecules can $strongly$
influence the molecular occupations, meanwhile the associating
light strength also can play a notable different role. In
particular, the novel effect of strong decaying molecular
oscillations (without or with molecular tunnelling in the deep- or
shallow-well case) can be revealed by adjusting these two
parameters. A simple physical picture for this effect is given by
comparing with the well-known two-color free-bound-bound PA case
[10], which clearly shows a promising way to create a stable
molecular condensate via coherent "double-well PA" technique.

\noindent

\begin{figure}[ht]
\includegraphics[width=0.6\columnwidth]{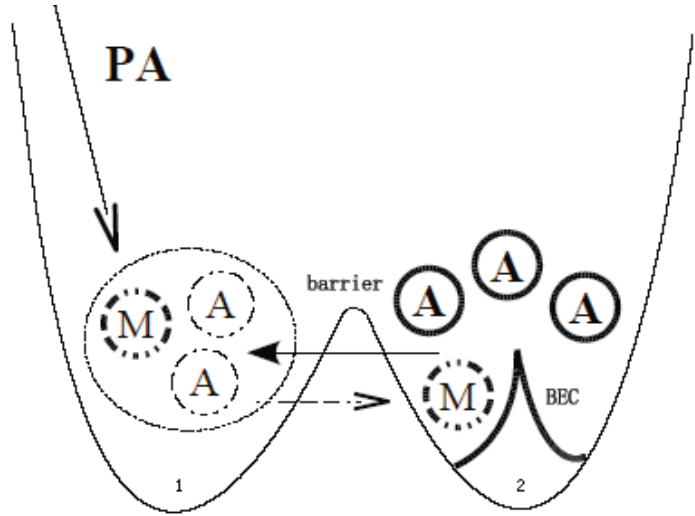}
\caption{A sketch of the coherent photo-association (PA) process
in a magnetic double-well potential. Initially, the atomic
component $A$ is entirely trapped in the right site, the coherent
PA only happens in the left site.} \label{1}
\end{figure}

Turning to the situation of Fig. 1, we assume for simplicity that
large number of Bose-condensed atoms are loaded into, say, the
right side of a double-well magnetic potential, and then the
Josephson tunnelling of particles may generate condensate in left
well which is initially vacuum. The atoms are photo-associated
into molecules while they tunnel as opposed to being
photo-associated when they have arrived in the left well. The
dynamics of double-well atomic condensates can be described by a
simple model [7] which includes the Kerr-type atomic interactions
occupying the same or different wells. However, the atomic
collisions effects on the dynamics of double-well condensates are
well-known [7, 11] and its strength can be tuned by the technique
of magnetic-field-induced Feshbach resonance [12], hence, to see
clearly the results of nonlinear PA interactions in presence of
particles tunnelling, we naturally ignore it for present purpose
(the strengths for the molecules or atom-molecule collisions are
yet not known [5]). In the second quantized notation, boson
annihilation operators for trapped atoms and molecules in two
wells are denoted by $a_1$, $a_2$, $b_1$ and $b_2$, respectively.
The free part of the total Hamiltonian are written as $H_0=-\Delta
a_1^\dagger a_1-\delta b_1^\dagger b_1$, where $\Delta$ and
$\delta$ are the magnetic and optical detunings respectively, and
generally, one should also consider the possible effects of these
detunings [5], but here we focus attention on the most interested
effects in this system due to the interactions of different modes
(i.e., the resonance case). Defining the optical Rabi coupling
frequency as $\gamma$, the four-mode Hamiltonian can be written as
$H=H_0+H_{int}$, where the interaction part is ($\hbar=1$)
\begin{equation}
H_{int}=-G_a(a_1^\dagger a_2+a_2^\dagger a_1)-G_b(b_1^\dagger
b_2+b_2^\dagger b_1)-\gamma({a_1^\dagger}^2b_1 +b_1^\dagger
a_1^2),
\end{equation}
and $G_a$ ($G_b$) is the atomic (molecular) tunnelling term, and
$\gamma$ represents an effective Rabi frequency characterizing the
coherent PA process applied for an $arbitrary$ time interval. For
simplicity, we have ignored the incoherent process of the
excited-state molecular damping or the effect of molecular
dissociating into those non-condensate atomic modes [13]. It is
important to realize that the tunnelling rate of the molecules is
exponentially suppressed compared to the tunnelling rate of the
atoms since the molecule is twice as heavy as the atom and the
tunnelling amplitude is proportional to $e^{-\sqrt{2m
V_0/\hbar^2}d}$, where $m$ is the mass, $V_0$ the barrier height,
and $d$ the barrier thickness. This means that the values of $G_a$
and $G_b$ are correlated instead of two independent parameters.
Obviously a conserved quantity exists for the present system:
$\sum_{i=1}^{2}{(a_i^\dagger a_i+2b_i^\dagger b_i)}\equiv N$,
where $N$ is the total atom number for a condensate of all atoms
or twice the total molecule number of a condensate of all
molecules.

Note that the main feature of our present scheme is that the
coherent PA process starts not directly in the initial trapped
atomic condensate, but in the "developing" atomic mode which is
initially in a vacuum state. The Heisenberg equations of motion
for the atomic and molecular modes read
\begin{eqnarray}
\dot{a}_1=iG_aa_2+2i\gamma a_1^\dagger
b_1,~~~\dot{a}_2=iG_aa_1; \nonumber\\
\dot{b}_1=iG_bb_2+i\gamma a_1^2,~~~\dot{b}_2=iG_bb_1.
\end{eqnarray}
The mean field approximation replaces the operators $a_i$ and
$b_i$ by $c$-numbers, i.e. $a_i\rightarrow \alpha_i$,
$b_i\rightarrow \beta_i$ ($i=1,~2$), then we have the following
simple $c$-number differential equations
\begin{eqnarray}
\dot{\alpha}_1=iG_a\alpha_2+2i\gamma \alpha_1^*
\beta_1,~~~\dot{\alpha}_2=iG_a\alpha_1; \nonumber\\
\dot{\beta}_1=iG_b\beta_2+i\gamma
\alpha_1^2,~~~\dot{\beta}_2=iG_b\beta_1,
\end{eqnarray}
where $\alpha_i$ and $\beta_i$ $(i=1,~2)$ are two complex numbers.
Such replacement is the so-called semiclassical approach analogous
to the Gross-Pitaevskii approximation used to describe an alkali
condensate [3, 4]. It is completely accessible to also incorporate
the molecular damping terms $\xi \beta_i$ phenomenologically into
the molecular-mode equations for our numerical study ($\xi$ is
proportional to the molecular damping rate), but we would not
consider it here for two physical reasons: firstly we mainly focus
on the role of coherent PA process in the double-well situation
and, comparing with the well-known single-well PA case, these
additional damping terms fail to bring any qualitatively new
results for the evolutions of $average$ molecular occupations in
our model; secondly, some techniques can be used to minimize the
effects of these quasi-bound molecular damping terms, such as the
Feshbach-resonance-assisted stimulated adiabatic passage (STIRAP)
method, proposed by Ling {\it et al.} through the MFA [14].

We remark that the MFA in our model should have a limited range of
validity. In particular, it certainly should break down and be
replaced by a fully quantized theory when the effects of quantum
$fluctuations$ become important. For example, by directly starting
from Eq. (2), we have analyzed the quantum dynamics and statistics
in the short-time and undepleted limits (without molecular
tunnelling) [9]. Of course, the precise criteria for the validity
of MFA in our model will be known only when the generalization to
the more interested long-time depleted case is available.
Nevertheless, one would expect that its range of validity should
include a certain variety of phenomena of interest, especially
when the concerned problem is the $average$ molecule numbers under
certain parameters [6]. In fact, the coherent two-color PA process
in an atomic BEC was already experimentally realized most recently
which can be well described by a mean-field three-mode model [4].
Interestingly, we will see that some nontrivial results really can
be revealed by the MF approach also in the present model.

To see concrete examples, we have analyzed the different dynamical
behaviors for the molecule numbers in two wells under the
conditions of deep- and shallow- well cases, respectively, and
compared these behaviors with that of single-well case, which
indicates a quite different but interesting way to create a stable
molecular condensate by using this double-well potential.

\noindent

\begin{figure}[ht]
\includegraphics[width=0.6\columnwidth]{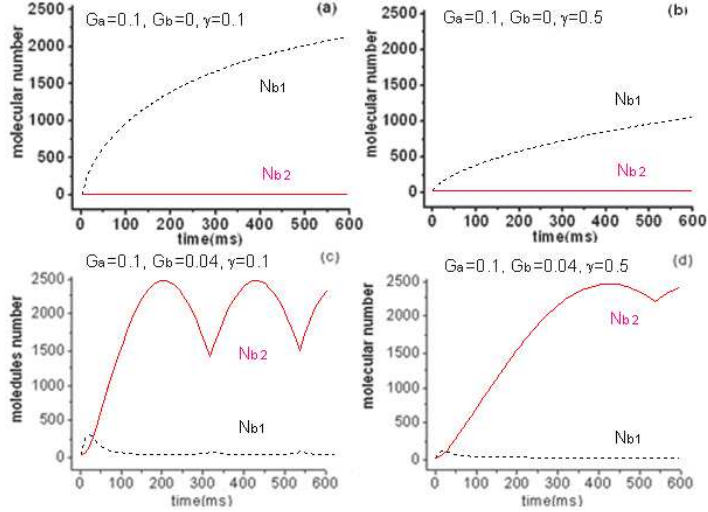}
\caption{(Color online) The molecular numbers in a deep
double-well case (a),(b) without or (c),(d) with molecular
tunnelling. The decaying molecular oscillations can be seen in
(c),(d) for larger PA strength. The two couplings (in units KHz)
and temporal length can be limited within validity range of MF
approach [6].} \label{2}
\end{figure}

\noindent
\begin{figure}[ht]
\includegraphics[width=0.6\columnwidth]{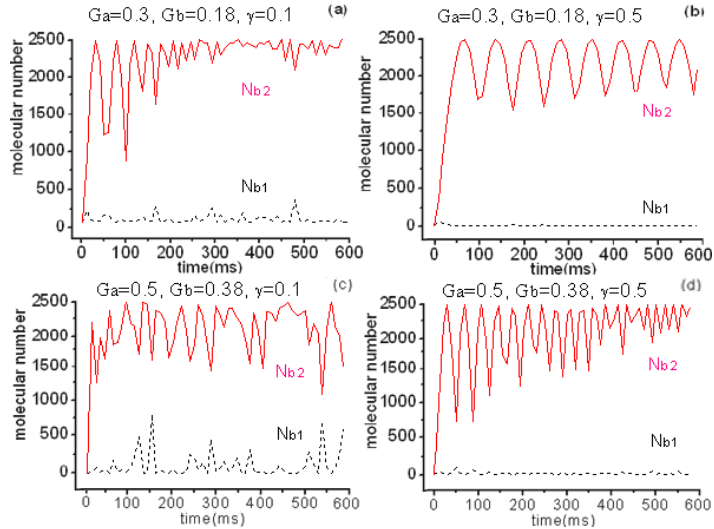}
\caption{(Color online) The molecular numbers in a shallow
double-well case with (a),(b) weak or (c),(d) strong molecular
tunnelling. Stronger couplings are not probed (for reducing valid
evolutions [6]). The tunnelling strength of molecules is dependent
on that of atoms. The decaying oscillations also can be seen in,
e.g., (d).} \label{3}
\end{figure}

$(i)~Deep$ $double$-$well$ $case.-$The results about the dynamical
evolutions of molecular numbers can be shown as Fig. 2 by
numerically solving Eq. (3). The initial atoms is typically $5000$
[5]. Clearly, if one only consider the atomic tunnelling, the PA
leads to an increasing of molecule number in left site, as it
should be. However, comparing with the single-well case (see the
later), the molecular formation which happens in an initial vacuum
instead of a trapped condensate will not need a strong PA light.
More interestingly, when we take the molecular tunnelling into
account, similar phenomena like that of two-color PA with
"molecular tunnelling" term [10] can be observed, i.e., almost
$all$ the formed molecules accumulate in $right$ site, meanwhile
the strong PA light leads to decaying molecular oscillations with
time. A similar physical picture can also be given for this [14]:
the $interference$ of two transitions (i.e., the transition of
left-well atoms to the molecules and then the transition of
left-well molecules to right-well molecules) under certain
parameters leads to its effectively zero occupation and then the
accumulations of photoassociated molecules in a different site.
Note that, for ensuring the validity range of MFA, the two
couplings and the temporal length are carefully reduced [6, 14].

$(ii)~Shallow$ $double$-$well$ $case.-$The evolutions of the
molecular numbers are shown in Fig. 3, which clearly shows the
different roles of atomic and molecular tunnellings in coherent
molecular formations. For much stronger atomic tunnelling and
strong PA light, we find strong decaying molecular oscillations
which, together with other useful techniques [5, 10, 14], may
provide a promising way to obtain more stable molecular condensate
via a double-well potential. Note that, this results are based on
our mean-field zero-dimensional model under the approximation of
density homogeneity within two wells, although it can be possible
to estimate the inhomogeneity effect in the Thomas-Fermi limit
[14].

\noindent
\begin{figure}[ht]
\includegraphics[width=0.6\columnwidth]{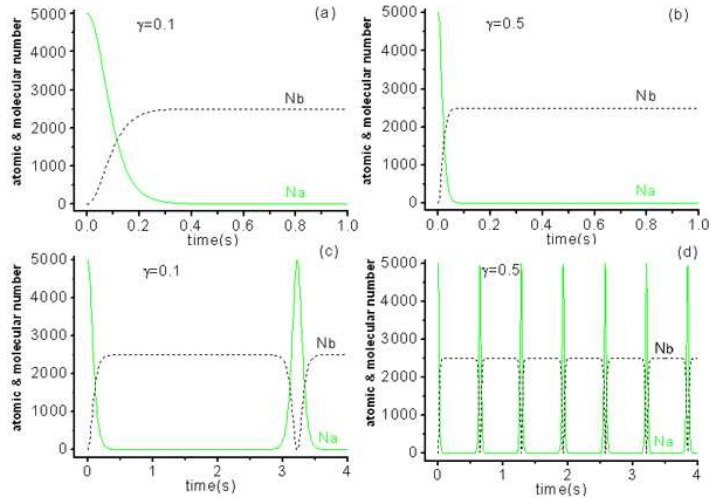}
\caption{(Color online) The populations of atoms
$N_a(=|\alpha|)^2$ and molecules $N_b(=|\beta|^2)$ in a single
well, for (a),(b) the short-time or (c),(d) the long-time case
(for the comparison with the double-well PA case).} \label{4}
\end{figure}

$(iii)~Single$-$well$ $case.-$For a clear comparison, we also plot
the short- and long-time behaviors of molecule numbers evolutions
for single well case as in Fig. 4(a)-(b), respectively. The $only$
control parameter now is the PA light strength. Obviously,
although stronger PA light induces rapid molecules increasing, it
also leads to rapid molecular oscillations for long evolution
time, which should be avoided in any actual experiment. Thereby
the considerable $decaying$ of molecular oscillations in our
scheme, especially for long evolution time, obtained by
controlling particle tunnellings and PA light strength, may shed
some new light on current experimental efforts towards a stable
molecular condensate.

We should emphasis again that, in our mean-field zero-dimensional
model, the only used approximation is the atomic density
homogeneity within two wells. Although we have ignored the atomic
collisions and the molecular damping as explained above, these
terms would be simple to include in the model especially in the
high coupling limits where their effects on the spatial structure
can be much reduced. Also it would be interesting to adopt the MFA
beyond zero-dimensional model [5, 15].

\bigskip

Summing up, we have numerically examined an interesting scheme for
creating a molecular condensate via coherent PA of the Josephson
tunnelling atoms in a double-well potential, focusing on the
$average$ molecular numbers in two wells. Since one of the main
problems of creating molecules by PA is that the same lasers also
tend to destroy the molecules and and thus it is of interest to be
able to move the molecules away from the lasers as quickly as
possible. Our scheme here may provide a promising way for creating
a stable molecular condensate via PA process in a double-well
potential. Based on a non-linear mean-field theory of a four mode
model, we found that for certain parameter regimes the molecules
accumulate in the well without the PA laser. We analyzed the
different roles of atomic and molecular tunnellings and that of PA
light strength in the molecular formations, through which an
interesting strong-decaying effect was observed for the right-site
molecular oscillations in the case of strong atomic tunnelling and
strong PA light strength (with PA process only happening in the
left site). This is quite different from the simple single-well
case. In fact our scheme can be viewed as further generalization
of two-color PA (only with "molecular tunnelling" [10, 14]).

Of course, even more important phenomena due to quantum
fluctuations of matter-wave fields, like the quantum statistics of
the resulting fields, can be investigated only by methods beyond
any mean-field approach. A powerful numerical technique based on
the $c$-number stochastic equations in the positive-$P$
representation of quantum optics [16] may be used to study these
intriguing subjects as in the extensive studies of two-color PA
process [5, 15], which may comprise the future works.
\bigskip

\noindent Note added: after the submission of this paper, we note
a very recent work of Olsen and Drummond in which a similar model
was presented and then solved by the methods of $c$-number
stochastic equations within, however, a quite different framework
of (coupled intracavity) optical down-converters [17].

\bigskip

\noindent We are grateful to the referees for their kind
suggestions which have lead to considerable improvement of this
paper. We also thank Shuai Kang, Jing Cheng and Yi-Jing Yan for
their helpful discussions. This work was supported by Wuhan Youth
Chenguang Project and the NSFC (10304020, 10574141).

\end{document}